\documentclass[aps,prl,twocolumn,nofootinbib,amsfonts,floatfix]{revtex4}
\usepackage{amssymb,graphicx}
\usepackage{amsmath}
\usepackage{hyperref} 
\usepackage{upgreek}

\def\e{{\rm e}}

\def\l{\left(}
\def\r{\right)}

\newcommand{\be}{\begin{equation}}
\newcommand{\ee}{\end{equation}}
\newcommand{\bea}{\begin{eqnarray}}
\newcommand{\eea}{\end{eqnarray}}
\newcommand{\bg}{\begin{gather}}
\newcommand{\eg}{\end{gather}}
\newcommand{\bseq}{\begin{subequations}}
\newcommand{\eseq}{\end{subequations}}

\renewcommand{\ln}{\mathop{\rm ln}\nolimits}

\begin{document}
\title{Evidence for a new particle on the worldsheet of the QCD flux tube 
  }

\author{Sergei Dubovsky${}^1$}
\author{Raphael Flauger${}^{1,2}$}
\author{Victor Gorbenko${}^1$}
\affiliation{${}^1$Center for Cosmology and Particle Physics, Department of Physics, New York University, New York, NY, 10003, USA}
\affiliation{${}^2$School of Natural Sciences, Institute for Advanced Study, Princeton, NJ 08540, USA}

\begin{abstract}
We propose a new approach for the calculation of the spectrum of excitations of QCD flux tubes. It relies on the fact that the worldsheet theory is integrable at low energies. With this approach, energy levels can be calculated for much shorter flux tubes than was previously possible, allowing for a quantitative comparison with existing lattice data. 
The improved theoretical control makes it manifest that existing lattice data provides strong evidence for a new pseudoscalar particle localized on the QCD fluxtube -- the worldsheet axion.
\end{abstract}

\maketitle

Recent advances in lattice simulations of quantum chromodynamics (QCD) have allowed to visualize confining strings quite vividly~\cite{Bissey:2006bz}\footnote{See {\tt http://www.physics.adelaide.edu.au/theory/staff\\/leinweber/VisualQCD/Nobel/} for animations.} and to measure the spectrum of their low lying excitations with impressive accuracy~\cite{Athenodorou:2010cs}. \\\indent
These lattice results have lead to an embarrassing situation for theorists. On the one hand, even for strings whose length is merely twice their width, many of the energy levels in the lattice simulations show remarkable agreement with the energy levels of a theory that the QCD string is certainly not described by, the bosonic string defined at the quantum level through light-cone quantization~\cite{Goddard:1973qh}. This Goddard--Goldstone--Rebbi--Thorn (GGRT) string is only Lorentz invariant in $D=26$. QCD flux tubes, however, originate from a relativistic theory in four dimensions and must be described by a worldsheet theory that respects Lorentz symmetry in $D=4$. The agreement is thus rather surprising.
On the other hand, existing theoretical techniques for calculating the flux tube spectra for the Lorentz-invariant Nambu--Goto (NG) string~\cite{Luscher:1980ac,Luscher:2004ib,Aharony:2010db} break down for the relatively short strings that can be simulated with current lattice techniques. 
To make matters more confusing, there is also a family of energy levels which disagree badly with the predictions made by the GGRT theory~\cite{Athenodorou:2010cs}. 

This {\it Letter} describes a theoretical framework to compute energy levels of the NG string for much shorter lengths than previously possible. Our better theoretical understanding allows us to explain both why there was agreement between QCD flux tubes and GGRT strings for many levels, and why there was disagreement for others. Rather interestingly, we show that the data implies the existence of a massive pseudoscalar resonance on the string worldsheet. The NG string itself is thus insufficient to describe QCD strings. We explain how to include this resonance into our framework and measure its mass and width from the data.

 Before presenting our results, let us first  describe the lattice data and compare it to the standard perturbative results~\cite{Luscher:1980ac,Luscher:2004ib,Aharony:2010db}. All numerical results we discuss are taken from~\cite{Athenodorou:2010cs} and are for gauge group $SU(3)$.
To measure the flux tube spectrum on the lattice, one calculates the discretized Yang--Mills\footnote{To avoid the string breaking due to quark pair production lattice simulations are performed in the pure glue theory. We refer to this theory as QCD.} partition function with a Wilson loop inserted at time $\tau=0$ and its conjugate at $\tau=T$, both winding around the compact spatial dimension with periodicity $R$. 
From the asymptotics of the partition function in the limit of large $T$, one deduces the energy of the ground state of a closed flux tube of length $R$. To measure the energies of excited states one deforms the shape of the Wilson line to project out lower-lying energy levels.
\begin{figure}[t!] 
 \begin{center}
 \includegraphics[width=3in]{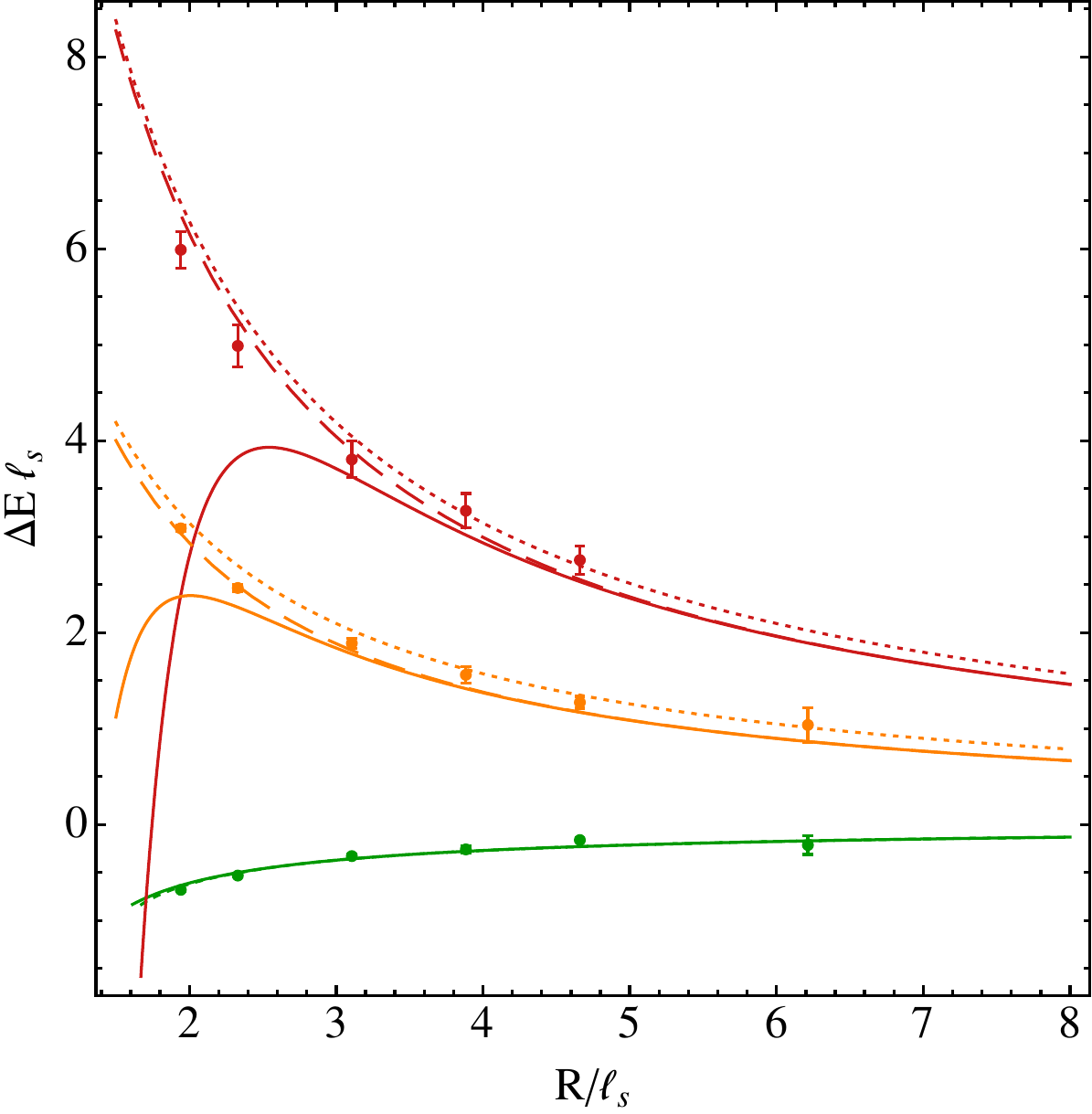}
 \caption{This plot shows $\Delta E=E-R/\ell_s^2$ as a function of the length of the flux tube $R$ for three different levels. The ground state is shown in green. States with spin 1 and with one and two units of longitudinal momentum are shown in orange and red, respectively. The value of $\ell_s$ was determined from the ground state data. The solid line shows the prediction of a derivative expansion. The dashed lines shows the prediction of the GGRT theory. For the spin 1 states the prediction for a free theory is shown as dotted lines.}
 \label{fig:gs}
 \end{center}
\end{figure}
The ground state energy as a function of the string length $R$ is shown in green in Fig.~\ref{fig:gs}. 
The data agrees remarkably well with the GGRT ground state energy.
Though surprising at first sight,  this agreement finds a straightforward explanation in an effective field theory approach. The excitations of a long QCD flux tube are Goldstone particles. They arise because the presence of a long, straight string spontaneously breaks the target space Poincar\'e symmetry $ISO(1,3)$ to $ISO(1,1)\times SO(2)$. For a recent discussion emphasizing this viewpoint, we refer the reader to~\cite{Dubovsky:2012sh}.
The standard method of calculating the effective string spectrum then is a derivative expansion, or equivalently an expansion in the small parameter $\ell_s/R$.
The non-linearly realized target space Lorentz symmetry imposes strong restrictions on the coefficients in this expansion and predicts all the coefficients up to $\ell_s^4/R^5$ \cite{Aharony:2010db}. These universal coefficients in the expansion are the same as those in the expansion of the GGRT ground state energy. Figure~\ref{fig:gs} illustrates that this universality alone is enough to explain the ground state data.

A bigger puzzle arises for excited states. 
Figure~\ref{fig:gs} shows the energy as a function of length for excited states with a single left-moving phonon with one and two units of longitudinal momentum. As before all the terms up to order $\ell_s^4/R^5$ in the $\ell_s/R$-expansion of the energies of these states are universal and coincide with the corresponding terms for the GGRT string. They agree well with the data at large $R$. However, for short strings the $\ell_s/R$-expansion  breaks down, and the universal terms no longer provide a good description of the data.

By itself the breakdown of the perturbative expansion for small $R$ is not surprising. One might, however, wonder why it works so much better for the ground state than for these excited states. A perhaps even more revealing question is why the data follows the GGRT energy spectrum so closely even in the regime where the low energy expansion breaks down. 
 
\begin{figure}[t]
 \begin{center}
 \includegraphics[width=2.8in]{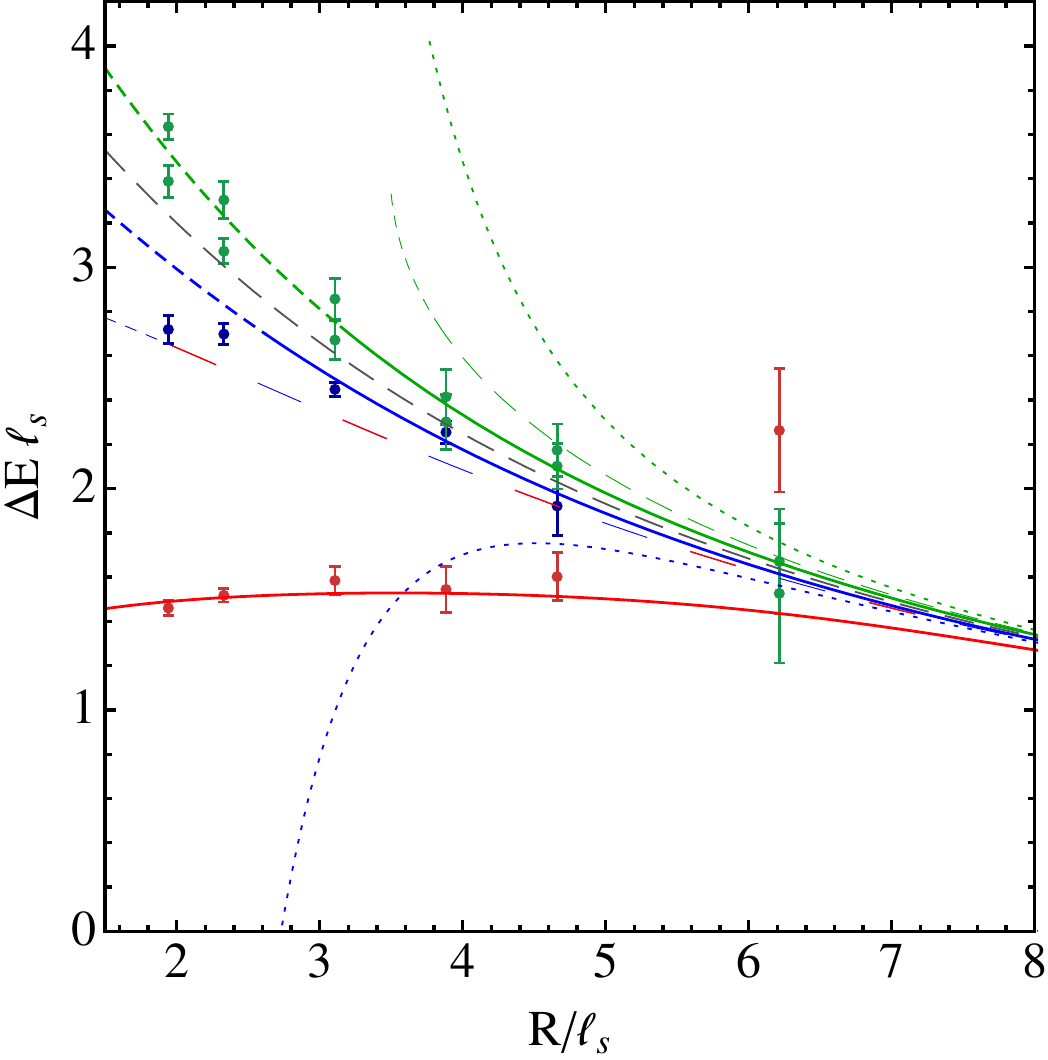} 
 \caption{This plot shows $\Delta E=E-R/\ell_s^2$ as a function of the length of the flux tube for the lowest lying states containing both left- and right-movers. The scalar and pseudoscalar states are shown in blue and red, respectively. The spin-2 states are shown in green. The solid lines show the theoretical predictions derived in the remainder of the paper, including the worldsheet axion in addition to the Nambu--Goto fields. The thinner, red/blue and green dashed lines show the prediction for the pseudo-scalar/scalar and tensor channel without the  axion. 
The dotted lines show the prediction of the $\ell_s/R$-expansion. The gray dashed line is the GGRT prediction. 
}
 \label{fig:e11tba}
 \end{center}
\end{figure}

The situation is similar, though slightly more intricate for excited states containing both a left- and a right-moving phonon each with one unit of longitudinal momentum. These two-particle states are conveniently characterized according to their transformation properties under the unbroken $SO(2)$. They break up into one scalar, one pseudoscalar (or equivalently an antisymmetric tensor), and two components of a symmetric traceless tensor.
The effective string theory predictions are again universal up to $\ell_s^4/R^5$ in the $\ell_s/R$-expansion. However, for these levels the universal terms of order $\ell_s^4/R^5$ for a relativistic string differ from those in the GGRT theory~\cite{Dubovsky:2012sh}. In the GGRT-theory, all these two-particle states are degenerate. For the relativistic string, representations with different spins are split. The splitting originates from the Polchinski--Strominger (PS) interaction \cite{Polchinski:1991ax,Aharony:2011ga,Dubovsky:2012sh} and is proportional to $D-26$. The numerical results of~\cite{Athenodorou:2010cs} are shown in Fig.~\ref{fig:e11tba} along with the predictions of the GGRT theory, the derivative expansion of the relativistic theory, and the theory predictions we will discuss later. The scalar and symmetric tensor levels follow the GGRT prediction rather closely at large radius, but for small radii a splitting is clearly visible. The splitting qualitatively agrees with the one predicted by the universal PS terms. However, the $\ell_s/R$-expansion breaks down at radii that are so large that a quantitative comparison is impossible. Even more noticeable is that the pseudoscalar state strongly deviates from the GGRT model.\footnote{One also notices a splitting between the two components of the symmetric tensor, but this splitting is due to lattice effects~\cite{Athenodorou:2010cs}.}
This data clearly calls for an alternative to the standard $\ell_s/R$ expansion.

Let us first identify the physical reason for the failure of the $\ell_s/R$ expansion for excited states.
To this end, it is instructive to first inspect the properties of the  $\ell_s/R$ expansion in the GGRT theory itself. Its exact spectrum is
\[
E_{{\small GGRT}}=\sqrt{{R^2\over\ell_s^4}+{4\pi^2(N-\tilde{N})^2\over R^2}+{4\pi \over \ell_s^2}\l N+\tilde{N}-{1\over 6}\r}\;.
\]
Here  $N$ and $\tilde{N}$ are the levels of an excited string state counting the amount of the longitudinal momentum carried by the left- and right-moving phonons separately, so that $2\pi(N-\tilde{N})/R$ is the total longitudinal momentum of the state. This formula immediately reveals a technical reason for the breakdown of the $\ell_s/R$ expansion. For excited states the expression under the square root not only involves powers of $\ell_s/R$, but also contains factors of $2\pi N$, which turn the $\ell_s/R$ expansion into a diverging asymptotic series even for relatively large values of $R$. 

To find a remedy let us reformulate the issue in more physical terms.
In general, the energy of an excited state is of the form
\[
E=\ell_s^{-1}{\cal E}(p_i\ell_s,\ell_s/R)
\]
where $p_i$ are the momenta of individual phonons comprising the state. These are quantized in units of $2\pi/R$ in a free theory, but we will see that this is modified in an interacting theory. So in reality the naive  $\ell_s/R$ expansion is a combination of two physically different expansions. The first expansion is an expansion in the softness of individual quanta compared to the string scale, $i.e.$ in $p_i\ell_s$.  The second expansion is a large volume expansion, $i.e.$ an expansion in $\ell_s/R$. 
 
 The key idea of our approach is to improve the convergence by disentangling the two expansions. 
To achieve this, we divide the calculation of the finite volume spectrum into two steps. We first calculate the (infinite volume) $S$-matrix. This can be done perturbatively provided the center of mass energy of the colliding phonons is small in string units. We will call this the momentum expansion. We then calculate the finite volume energies from this $S$-matrix. Conceptually, this step is not related to the perturbative expansion used to calculate the $S$-matrix. However, a prescription for this step for a given $S$-matrix only exists in two cases. First, there is a perturbative procedure due to L\"uscher \cite{ Luscher:1985dn,Luscher:1986pf}  which is routinely used in lattice calculations and applicable for states with energies below the inelastic threshold. Second, for two-dimensional integrable theories, there is the Thermodynamic Bethe Ansatz (TBA)~\cite{Zamolodchikov:1989cf,Dorey:1996re}, a set of integral equations, whose solution yields the exact spectrum of the theory on a circle from the $S$-matrix. 

In our case the particles are massless and we cannot rely on L\"uscher's method. However, let us ignore this for now and give the prescription to extract the finite volume spectrum from the $S$ matrix in a way that allows us to discuss both methods in parallel.\footnote{For simplicity, we suppress flavor indices and present the equations in somewhat condensed form. For more details see equations (27)-(32) in reference~\cite{Dubovsky:2012wk}} For massless particles, it is convenient to choose $p_i$ to be positive and to divide particles into left- and right-movers.
The energy of the state then takes the form
\be
\label{TBAenergy}
\Delta E=\sum_i p_i+W_E\;,
\ee
where $W_E$ represents winding effects from virtual quanta traveling around the circle.
The particle momenta satisfy a modified periodicity condition,
\begin{gather}
\label{TBAperiodicity}
\epsilon(p_{i})=2\pi n_{i}/R\;,
\end{gather}
 where
 \begin{gather}
 \label{TBAeq}
\epsilon(p)=  p+\frac1R\sum_{j}2\delta(p,p_j)+W_P\;,
\end{gather}
and $n_i$ are positive integers. Interactions thus modify the quantization condition for momenta in two ways. First, real particles scatter with each other explaining the infinite volume scattering phase shift $2\delta(p_i,p_j)$ in Eq.~(\ref{TBAeq}). Second, it is modified by winding corrections represented by $W_P$. 

In L\"uscher's approach the winding corrections $W_E$, $W_P$ are calculated perturbatively. For theories with a mass gap $\mu$ they are exponentially suppressed as $\e^{-\mu R}$. In analyses of lattice results it is thus common to use the massive version of equations (\ref{TBAenergy}), (\ref{TBAperiodicity}), with $W _{E,P}=0$. In the context of integrable field theories, the resulting simplified equations are known as the asymptotic Bethe Ansatz.

In massless theories winding corrections are only power-law suppressed, requiring us to work harder and to use insights from integrable theories. 
%For a generic $S$-matrix, even writing the excited state TBA equations in closed form can be challenging, and a certain amount of guesswork is typically involved. 
The exact form of the excited state TBA equations for the GGRT theory is known explicitly \cite{Dubovsky:2012wk} 
\begin{gather}
\label{WEgen}
W_E={1\over \pi}\sum_{l,r}\int_0^\infty dp' f_{l(r)}(p')\,,
\\
W_{P}^{l(r)}(p)={1\over \pi R}\int_0^\infty dp' {d2\delta(p,p')\over dp'}f_{r(l)}(p')\;,
\nonumber
\end{gather}
where the $l(r)$ subscript refers to left(right)-movers, and the densities $f_{l(r)}(p)$ are reminiscent of a thermal bath at temperature $1/R$,
\be
\label{thermal}
f_{l(r)}(p)=\ln\l 1-\e^{-R\epsilon_{l(r)}(p)}\r\,.
\ee
The pseudo-energies $\epsilon_{l(r)}(p)$ are then determined by solving the integral TBA equations (\ref{TBAeq}). Note that also for the integrable sinh-Gordon theory
the TBA equations take the same form  \cite{Teschner:2007ng}, which strongly suggests that this form is universal for purely elastic scattering.

In the GGRT theory the phase shift takes the simple form
 \be
 \label{GGRTphase}
2\delta_{GGRT}(p_l,p_r)=\ell_s^2 p_lp_r\;,
\ee
allowing for an exact solution of the TBA equations~\cite{Dubovsky:2012wk}.
For  future use note, that for a state with a pair of left- and right-movers with equal and opposite
momenta $\hat{p}$,  
 this phase shift together with Eqs. (\ref{TBAeq}), (\ref{WEgen}) and (\ref{thermal}) results in the linear dispersion
relation for pseudo-particles
\be
\label{GGRTe}
\epsilon_{GGRT}(p)=c p
\ee
where $c$ is a solution (the one which approaches the free theory value $c=1$ at $\ell_s\to 0$) of the following quadratic equation
\be
\label{cGGRT}
c=1+\ell_s^2{\hat p\over R}-{\pi \ell_s^2\over 6 R^2c}\;.
\ee
This translates into
\be
\label{WPE}
W^{GGRT}_P(\hat{p})=-{\pi\ell_s^2\hat{p}\over 6 R^2c}\;,\;\;\;\; W^{GGRT}_E=-{\pi\over 3 Rc}\;.
\ee

The worldsheet theory of a QCD flux tube is not integrable and the situation seems hopeless. However, as a consequence of the target space translation symmetry the theory is weakly coupled at low energies and the low energy scattering is dominated by purely elastic processes. Furthermore, the low-energy scattering amplitudes agree with those of the GGRT theory, which is integrable and for which we do know the exact form of TBA equations and winding corrections. We can thus use  the TBA equations for the GGRT theory as a zeroth order approximation and incorporate higher order contributions in the momentum expansion as corrections to the scattering phase shift. In principle, this should be done both in asymptotic and winding parts of the TBA equations. In practice, we include corrections only in the asymptotic part and use the GGRT phase shift
and pseudo-energy (\ref{GGRTe})  for windings. This is justified by the form of Eq.~(\ref{thermal}). It ensures that the winding corrections receive their dominant contributions from virtual quanta with momenta below $1/R$, much softer than the real quanta. It is also important to note that the momenta of the real quanta for multiparticle states containing both left- and right-moving phonons are softer than the free theory estimate $2\pi n_i/R$. This is immediate from (\ref{TBAperiodicity}),(\ref{TBAeq}) and the fact that the GGRT phase shift (\ref{GGRTphase}) grows in the UV.
This indicates that one should expect a better than naive agreement between the QCD flux tube and the GGRT spectrum. As soon as the relevant momenta are small enough, the two theories have similar infinite volume $S$-matrices and as a result should have similar finite volume spectra. This fact is lost in the  conventional perturbative expansion.

Let us first apply this logic to the purely left-moving states. For these states the asymptotic Bethe Ansatz is trivial. The GGRT winding corrections are small, and one expects the spectrum to be close to that of a free theory. The dotted line in Fig.~\ref{fig:gs} shows that this expectation is correct. This eliminates the mystery for these states.

For states containing both left- and right-movers, we need to take into account corrections to the GGRT phase. The leading one-loop correction to the amplitude is universal and takes the PS form \cite{Dubovsky:2012sh}
\be\label{PSphase}
2\delta_{PS}(p_l,p_r)=\pm {11\ell_s^4\over 12\pi}(p_lp_r)^2\;,
\ee
where $``+"$ refers to the scalar and pseudoscalar channels and $``-"$ to the symmetric tensor channel.
\begin{figure}[t]
 \begin{center}
 \includegraphics[trim=4.5cm 3.5cm 5cm 3.5cm,width=2.4in]{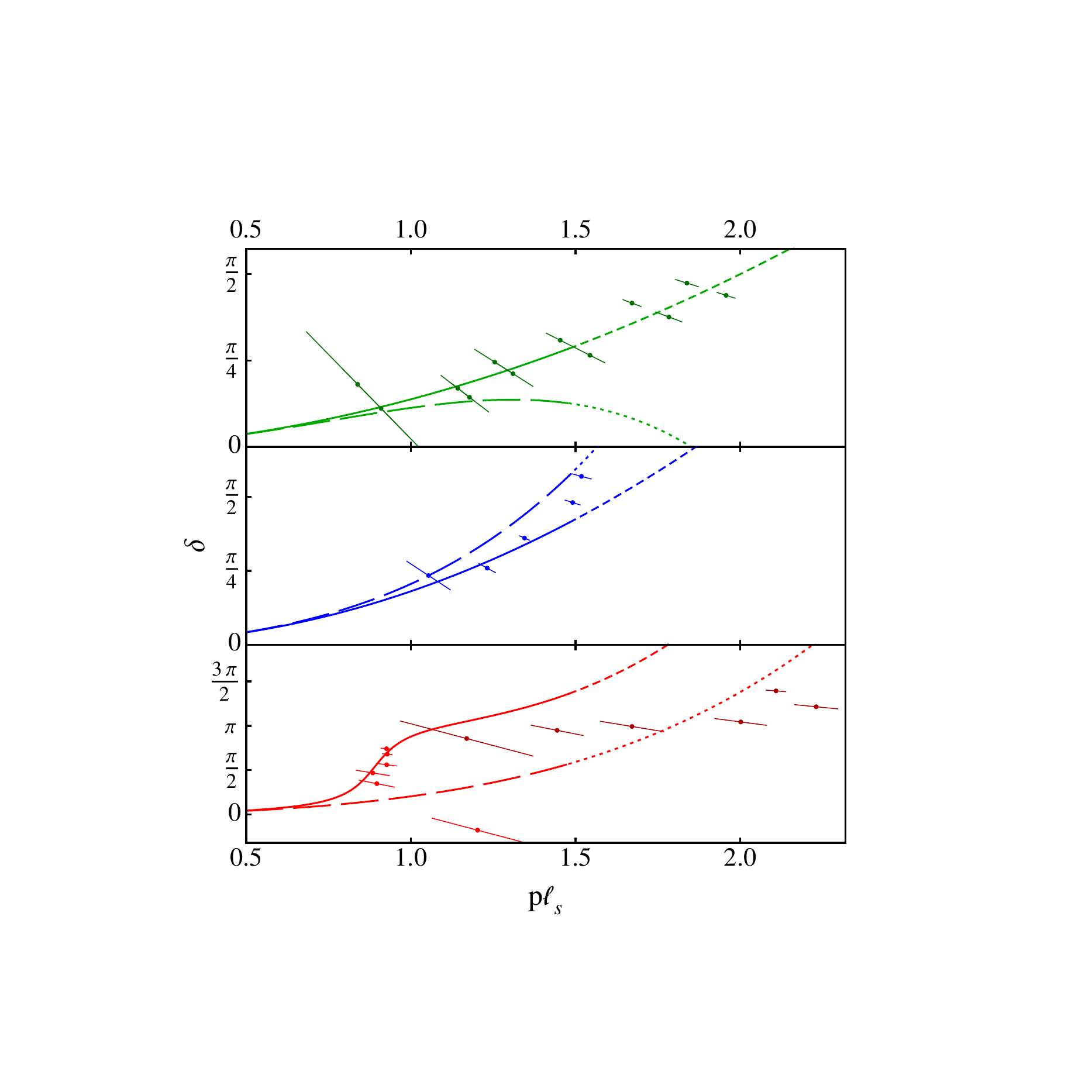} 
 \caption{This plot shows the scattering phase shift $\delta$ for two Goldstone bosons as a function of the center of mass momentum in the symmetric traceless, scalar, and antisymmetric channel in the top, middle, and bottom panel, respectively. The solid and the long dashed lines show the theoretical prediction with and without the worldsheet axion, respectively.}
 \label{fig:resonance}
 \end{center}
\end{figure}

Including this contribution in the asymptotic Bethe Ansatz results in a significant improvement for the scalar and symmetric tensor levels as shown by the thin dashed lines in Figure~\ref{fig:e11tba}.\footnote{The lines are dashed where~\eqref{PSphase} becomes comparable to~\eqref{GGRTphase} and our approximations are unreliable.} Notice that at this point we have not introduced any new parameters.
The improved theoretical control makes it manifest that the anomalous behavior of the pseudoscalar level cannot be blamed on the breakdown of the perturbative expansion and a qualitatively new ingredient is needed.
The energy of the anomalous level is roughly independent of the radius. This suggests that the most straightforward way to explain this level is the introduction of a massive pseudoscalar particle $\phi$ on the worldsheet. The leading interaction compatible with non-linearly realized Lorentz invariance for such a particle is a coupling to the topological invariant known as the self-intersection number of the string worldsheet
\be
\label{axionint}
S_{int}={\alpha\over  8\pi }\int d^2\sigma \phi K^i_{\alpha\gamma}K^{j\gamma}_\beta\epsilon^{\alpha\beta}\epsilon_{ij}\,,
\ee
where $K^i_{\alpha\gamma}$ is the extrinsic curvature of the worldsheet.
The existence of this worldsheet $\theta$-term for a string in a four-dimensional target space was pointed out by Polyakov \cite{Polyakov:1986cs}, and it was suggested that it should be generated on the flux tube worldsheet in the presence of the bulk $\theta$-term \cite{Mazur:1986nr}. Given this coupling, it is natural to refer to the field $\phi$ as the worldsheet axion.

The axion appears as a resonance in the scattering of Goldstone bosons with antisymmetric flavor wave function and it also contributes to the scattering in the scalar and symmetric tensor channels through $t$- and $u$-channel diagrams. It is thus readily included in the TBA equations.
By following the strategy outlined above, i.e. by making use of the GGRT expressions (\ref{GGRTe}), (\ref{cGGRT}) and (\ref{WPE}) for winding corrections, we arrive at the following modified quantization condition
\be
\label{LRper}
c \hat{p} R+2\delta_{PS}+2\delta_{res}=2\pi\;,
\ee
where 
\[
2\delta_{res}=\sigma_1{\alpha^2\ell_s^4\hat{p}^6\over 8\pi^2(4\hat{p}^2+m^2)}+2\sigma_2 \tan^{-1}\l{\alpha^2\ell_s^4\hat{p}^6\over 8\pi^2 (m^2-4\hat{p}^2)}\r
\]
is the axion contribution to the phase shift as derived from (\ref{axionint}) with $\sigma_1=(-1,1,1)$, $\sigma_2=(0,0,1)$,
 for scalar, symmetric and antisymmetric channels correspondingly. By solving the periodicity condition (\ref{LRper}) and plugging the result in (\ref{TBAenergy}) with the GGRT expression (\ref{WPE}) for the winding correction $W_E$ we 
 arrive at the final result for the energies.

 By fitting the two free parameters (the axion mass $m$ and the coupling $\alpha$) to the data, we obtained the spectrum shown as solid lines in Fig.~\ref{fig:e11tba},  which corresponds to $m\ell_s=1.85^{+0.02}_{-0.03}$, confirming the heuristic analysis of~\cite{Athenodorou:2010cs}, and $\alpha=9.6\pm 0.1$. The error bars represent the statistical uncertainty only. Based on a comparison of the two symmetric tensor levels and a comparison of the states with zero and one unit of longitudinal momentum, we estimate the systematic and theoretical uncertainties to be about a factor of five larger. 

Note that unlike the heuristic formulae of ~\cite{Athenodorou:2010cs}, designed to fit the pseudoscalar channel only, the TBA analysis predicts also the energy shifts in the scalar and symmetric tensor channels associated with the same resonance. As seen in Fig.~\ref{fig:e11tba}
these shifts result in a significantly better agreement with the data. 

Further support for existence of this axion comes from data for the next excited level in the pseudoscalar channel. We reverse the logic and use the TBA equations to determine the scattering phase shift from the finite volume spectrum, which is the standard approach in lattice QCD. The resulting phase shifts for the pseudoscalar, scalar and symmetric tensor channel are shown in Fig.~\ref{fig:resonance}. For the pseudoscalar, it exhibits a characteristic resonance shape with the phase shift crossing $\pi/2$. The phase shift extracted from the data for the pseudoscalar state we discussed so far is shown in light red. The dark red points show the phase shift extracted from the data for the next excited pseudoscalar state also taken from~\cite{Athenodorou:2010cs}. Let us stress that presenting the data this way provides a very convincing case for the existence of a pseudoscalar resonance 
(by the very definition of what a resonance is), without relying on any fitting procedure.
%Somewhat suggestively, two data points in the scalar channel near $\delta=\pi/2$ also deviate from the theoretical expectation. This might indicate the presence of a scalar resonance with $m\approx3/\ell_s$. Its existence should then be more easily detectable in excited states with the same quantum numbers. 

In summary, the TBA approach provides better theoretical control over flux tube spectra than the standard perturbative expansion. Presently, this is the only available method for calculating the 
spectrum of flux tube excitations for the flux tube lengths probed on the lattice.
We conclude that existing lattice data provides strong evidence for the existence of a new particle -- the worldsheet axion. 
In a forthcoming publication~\cite{TBA} we will present the details of our analysis and elaborate on the diagrammatic interpretation of the TBA method. 
We will also show the evidence for the same resonance in the lattice data for more highly excited states.

We thank  Daniele Dorigoni, Giga Gabadadze, Joe Polchinski,  Matt Roberts, Arkady Tseytlin, Gabriele Veneziano, and Konstantin Zarembo for useful discussions. This work is supported in part by the NSF grant PHY-1068438. The work of R.F. has been supported in part by the National Science Foundation under Grant No. NSF-PHY-0855425 and NSF-PHY-0645435.

\bibliographystyle{h-physrev3}
\bibliography{dlrrefs}
\end{document}